\documentclass[iicol,pdflatex,sn-mathphys-num]{sn-jnl}

\usepackage{graphicx}%
\usepackage{multirow}%
\usepackage{amsmath,amssymb,amsfonts}%
\usepackage{amsthm}%
\usepackage{mathrsfs}%
\usepackage[title]{appendix}%
\usepackage{xcolor}%
\usepackage{textcomp}%
\usepackage{manyfoot}%
\usepackage{booktabs}%
\usepackage{algorithm}%
\usepackage{algorithmicx}%
\usepackage{algpseudocode}%
\usepackage{listings}%

\theoremstyle{thmstyleone}%
\theoremstyle{thmstyletwo}%

\theoremstyle{thmstylethree}%

\begin{document}

\title[An Active Contour Model for Silhouette Vectorization]{An Active Contour Model for Silhouette Vectorization using B\'{e}zier Curves}


\author*[1]{\fnm{Luis} \sur{Alvarez}}\email{lalvarez@ulpgc.es}

\author[2]{\fnm{Jean-Michel} \sur{Morel}}\email{jeamorel@cityu.edu.hk}
\equalcont{These authors contributed equally to this work.}

\affil*[1]{\orgdiv{Departamento de Informática y Sistemas}, \orgname{Universidad de Las Palmas de Gran Canaria}, \orgaddress{\street{Campus de Tafira}, \city{Las Palmas de G.C.}, \postcode{35017}, \country{Spain}}}

\affil[2]{\orgdiv{Department of Mathematics}, \orgname{City University of Hong Kong}, \orgaddress{\street{Tat Chee Avenue, Kowloon Tong},  \country{Hong Kong}}}


\abstract{In this paper, we propose an active contour model for silhouette vectorization using cubic B\'ezier curves. Among the end points of the B\'ezier curves, we distinguish between corner and regular points where the orientation of the tangent vector is prescribed. By minimizing the distance of the B\'{e}zier curves to the silhouette boundary, the active contour model optimizes the location of the B\'{e}zier curves end points, the orientation of the tangent vectors in the regular points, and the estimation of the B\'{e}zier curve parameters. This active contour model can use the silhouette vectorization obtained by any method as an initial guess. The proposed method significantly reduces the average distance between the silhouette boundary and its vectorization obtained by the world-class graphic software Inkscape, Adobe Illustrator, and a curvature-based vectorization method, which we introduce for comparison. Our method also allows us to impose additional regularity on the B\'{e}zier curves by reducing their lengths.}




\maketitle

\section{Introduction}

Most digital images are acquired as a raster image, defined by values on a regular grid of pixels. When the image is binary, the shapes that are observed in the image appear as unions of square pixels. This pixel representation, viewed as a subset of the plane, has a very weak structure. For example, most boundary pixels add artificial corners to the shape, a phenomenon known as the pixelation effect \cite{bachmann2016perception}. The shape cannot be directly rescaled, translated, rotated, etc. without accentuating these defects\cite{attneave1957physical,elder1999edges,torralba2009many}.
Silhouette vectorization (also called silhouette tracing) is a process in which the outline or silhouette of an object in an image is extracted and represented as a vector graphic. Vectorization involves converting binary raster shapes, which are simply unions of pixels, into a vector graphics representation where the shape boundary is described by mathematical equations (usually cubic B\'{e}zier curves). Silhouette vectorization is particularly useful in various applications, including graphic design, computer vision, and image processing. The main steps involved in silhouette vectorization are usually: 

\begin{enumerate}
    \item Decomposition of the silhouette boundary as a collection of closed curves.
    \item For each closed curve, $C(t)$, selection of a subset of points $\{C(t_n)=\bar x_n \}$ that are the end points of the cubic B\'{e}zier curves used to approximate the silhouette boundary. 
    \item Estimation of the parameters of the cubic B\'{e}zier curves approximating $C(t)$.  
\end{enumerate}
Among the  B\'{e}zier curve end points we can  distinguish between corner points, where the boundary curves ending at the point change direction, and regular points, where the curves have a prescribed tangent vector orientation. We denote by $Corners \subset \{t_{n}\}$, the set of corner points. 
In this paper, we propose an active contour model that fits the cubic B\'{e}zier curves to $C(t)$ by optimizing the choice of the interpolating values $t_n$, the orientation of the tangent vectors at the regular points, and the B\'{e}zier curve parameters. This active contour model is based on minimization of the distance of the B\'{e}zier curves to the silhouette boundary. The main contributions of the paper are: 

\begin{enumerate}
    \item The formulation of the active contour model optimizing the approximation of the silhouette boundary by a collection of cubic B\'{e}zier curves. 
    \item The representation, in the active contour model, of the control points of the B\'{e}zier curves in terms of the tangent vectors at the end points, which makes it easy to constrain the value of the tangent vectors at the regular points. 
    \item The active contour model optimizes not only the position of the control points of the B\'{e}zier curves, but also the positions of their end points and the orientation of the tangent vectors at the regular points. 
    \item The method can be used to improve the results obtained by any vectorization method. In particular, we show that it significantly reduces the average distance between the silhouette boundary and its vectorization obtained by Inkscape, Adobe Illustrator and a basic curvature-based vectorization method that we introduce for comparison purposes. 
    
    \item  The proposed active contour model also allows for imposing additional regularity on the B\'{e}zier curves to remove, if needed, undesirable irregularities. 
\end{enumerate}

\section{Related work }

\paragraph{Image vectorization and its advantages}
Image vectorization is a technique to transform a raster image into an image that avoids pixelation defects by representing the image by a finite number of smooth curves that separate regions endowed with a constant or parametric color model. This \textit{vector graphic} representation is highly successful~\cite{tian2022survey,dziuba2023image}. Image vectorization gives a grid-invariant representation~\cite{he2021finding,he2022silhouette,he2023topology}, enabling content editing~\cite{orzan2008diffusion}, the reconstruction of line drawings and cartoons~\cite{hilaire2006robust,noris2013topology,favreau2016fidelity,zhang2009vectorizing}, and clipart \cite{yang2015effective,favreau2017photo2clipart}. In most vectorized images ~\cite{he2023binary,he2023viva,reddy2021im2vec,li2020differentiable} the image regions are encoded by a parametric color description, often simply a constant color in each region, separated by curves composed of B\'{e}zier curves. These parametric curves have precise grid-independent shapes that justify their name of Scalable Vector Graphic (SVG)~\cite{ferraiolo2000scalable}.When displayed on a screen or printed, such images must be converted back to raster images. This rendering process reverses the vectorization process and simulates optical blur using so-called anti-aliasing filters.  All graphic software ~\cite{AI} enables this instant conversion at any required image scale. Note that more advanced vector graphic representations than SVG have been proposed involving mesh generation~\cite{sun2007image}  and elliptic equations used as interpolators ~\cite{orzan2008diffusion}. Recent deep learning methods that optimize the shape and color control points of deformable objects have also been proposed~\cite{lopes2019learned,li2020differentiable,vinker2022clipasso,ma2022towards}. 
In the context of silhouette vectorization, cubic B\'{e}zier curves are often used to represent smooth and precise curves defining the silhouette boundary. The control points of the B\'{e}zier curves are adjusted to match the shape of the extracted contours. There exist many different ways to fit the curve $C$ by cubic B\'ezier curves (see, for instance, 
\cite{pastva1998bezier}).  
The open source Inkscape \cite{Selinger2003PotraceA} trace bitmap tool performs silhouette vectorization based on a polygonal approximation of the silhouette boundary and cubic B\'{e}zier curves. In \cite{CINQUE1998821}, the authors use B\'{e}zier curves including a parameter controlling the complexity and resolution of the approximation process. In \cite{PlSt83} the authors use B\'{e}zier curves with tangent vector specifications. In \cite{HHM22} the authors use the affine scale-space to select the end points of the B\'{e}zier curves. In \cite{HHM23} the authors propose a method for vectorization of color images.

\paragraph{Variational models of shape approximation, curve shortening models}
Starting as early as 1972 variational image segmentation \cite{pavlidis1972segmentation} models led to  the general \textit{Mumford-Shah functional}~\cite{mumford1989optimal,morel2012variational}:
\begin{multline} \label{equ:MS}
 \mathcal{E}_f(u,\Gamma)=\int_{\Omega \setminus \Gamma}\|u(x)-f(x)\|_2^2\,dx \ + \\ 
 \int_{\Omega\setminus \Gamma}\|\nabla u(x)\|_2^2\,dx+\lambda \int_{\Gamma}\,d\sigma(x)\;.
\end{multline}
The reconstructed image $u:\Omega\to\mathbb{R}^d$ ($d=1$ or $3$) is obtained as an approximation of the input image $f:\Omega\to\mathbb{R}^{d}$. In~\eqref{equ:MS}, $\Gamma\subset\mathbb{R}^2$ is the discontinuity set of $u$, $\lambda>0$ is the regularization parameter, and $d\sigma$ is the length element. The pair $(u, \Gamma)$ can be understood as a set of smooth regions whose smoothness is controlled in the second term, separated by curves whose smoothness is enforced by minimizing the length in the third term. Variational segmentation models such as~\eqref{equ:MS} are non-convex and non-smooth optimization problems~\cite{kornprobst2006mathematical,cai2013two} that require a sophisticated optimization machinery~\cite{pock2009algorithm,chan2001level,ambrosio1990approximation}. The application to the approximation of a single shape in an image is the celebrated snakes model \cite{kass1988snakes} which introduced the notion of active contour models minimizing again a functional where the smoothness of the contour competes with its fidelity to the image's discontinuities. The complexity of its optimization has led to a rich literature \cite{xu1998snakes,amini1988using,bresson2007fast,caselles1997geodesic,cao2003geometric,amini1988using}.

\paragraph{Curve smoothing and curve shortening}Bounded  shapes in the plane can be modeled as Caccioppoli sets \cite{miranda2003caccioppoli}. Their boundary is a union of nested Jordan curves \cite{ambrosio2001connected}. When this boundary is a simple curve, we call the shape a silhouette. Smoothing the shape then amounts to smoothing this simple curve. The curve shortening flow, or evolution of a simple curve by the intrinsic heat equation, was shown in \cite{grayson1987heat} to be a consistent model that progressively reduces the complexity of the curve until it collapses at a round point \cite{chou2001curve}. It can be viewed as the unconstrained gradient descent of the third term of the Mumford-Shah functional. As anticipated in \cite{asada1986curvature}, the analysis of a curve focuses on its extrema of curvature \cite{do2016differential}. The curve shortening flow is often interpreted as a \textit{motion by curvature} \cite{osher1988fronts} because, in the shortening flow, the points of the curve move with a velocity proportional to their curvature. An important variant is the affine curve shortening flow, which has very similar properties, but also commutes with affine perspective deformations, a desirable property in shape analysis \cite{sapiro1993affine,tannenbaum1997invariant,angenent1998affine}. There is a very fast consistent scheme \cite{moisan1998affine} for this curve evolution, which is used in \cite{he2021accurate,he2023binary} for shape vectorization. The method~\cite{he2022vectorizing} develops an extension of the silhouette affine shortening to general image vectorization, and a variant~\cite{he2023topology} focuses on evolving the curves that support T-junctions.The affine curve shortening flow has also been shown to be very efficient for the multiscale detection of curve corners \cite{alvarez1997affine,alvarez2017corner} and of multiple junctions. See also \cite{magni2013motion} for the curvature motion of planar graphs.

In the seminal paper \cite{CKS97}, Caselles et al. introduced the geodesic active contour formulation given by 
\begin{equation}
\mathcal{E}(B)\equiv\int_{0}^{1}g(B(s))\Vert B'(s)\Vert ds, \label{eq:energy_caselles}%
\end{equation}
where $B:[0,1]\rightarrow R^2$. $\mathcal{E}(B)$ corresponds to a curve length
definition obtained by weighting the Euclidean element of length
$\Vert B'(s)\Vert ds$ by $g(B(s)).$ This simple and elegant formulation has been successfully applied to many different scenarios. Geodesic contours  model with prior shapes information have been formulated in \cite{CTT02} and \cite{LCHS05} using a level set evolution approach where the evolving curve is represented as the boundary of  the zero level set of an evolving surface.  In \cite{Al3Dc22}, it has also been applied to the regularization of 3D curves using as function $g(\bar x)$: 
\begin{equation}
    g(\bar x)=d_C(\bar x)+w,
    \label{eq:g}
\end{equation}
where $d_C(\bar x)$ is the Euclidean distance of any 3D point $\bar x$ to a prescribed curve $C:[0,T]\rightarrow R^3$ and $w \geq 0$. For this choice of $g(\bar x)$, the minimization of (\ref{eq:energy_caselles}) yields a curve close to $C(t)$. The parameter $w$ is a regularization factor that penalizes large values of the length of $B(s)$ in the energy.

The active contour model that we propose in this paper is based on the geodesic active contour (\ref{eq:energy_caselles}) with $g(\bar x)$ given by (\ref{eq:g}) (but in $R^2$ and with $C(t)$ a closed curve on the silhouette boundary). In our case, the curve $B(s)$ is parameterized using a collection of cubic B\'{e}zier curves with prescribed tangent orientation in the regular points. Parameterized active contours have been used, for instance, for the detection and tracking of circles \cite{CGA18}, and ellipses \cite{AGCTTC22}. Using parameterized active contours simplifies the minimization of energy (\ref{eq:energy_caselles}) and, in general, we can use $w=0$ as a regularization parameter, because the parameterized curves used are, in general, smooth.

\section{The active contour model}

The input of the active contour model we propose is given by : 

\begin{enumerate}
    \item A closed curve $C:[0,T]\rightarrow R^2$.
    \item An initial  collection of interpolating values $\{t_{n}%
\}_{n=1,..,N}$ with $0\leq t_{1}<t_{2}<\cdot$ $\cdot\cdot<t_{N}<T$, with the associated 2D points $\bar x_n=C(t_n)$. As we deal with closed curves, to simplify
the presentation, we consider that the position $N+1$ is identified with the
first position, that is $t_{N+1}=t_{1}.$ For any
$t,t^{\prime}\in\lbrack0,T]$ with $t^{\prime}<t,$ we consider that
$[t,t^{\prime}]=[t,T]\cup\lbrack0,t^{\prime}]$. We denote by $L_n$ the length of $C(t)$ between $t_n$ and $t_{n+1}$. 
   \item The set of corner points $Corners\subset \{t_{n}\}$ where no restriction is imposed about the B\'{e}zier curve  parameters. 
   \item $\mathcal{T}(\alpha_{n})=(\cos(\alpha_{n}),\sin(\alpha_{n}))^{T} $ the initial unit
tangent vectors to $C(t)$ at $t_n$ 
and $\mathcal{T}(\alpha_{n})^{\perp}=(-\sin(\alpha_{n}),\cos(\alpha_{n}))^{T}$
their orthogonal direction. In the regular points, $t_n \notin Corners$, we constrain the tangent vector of the B\'{e}zier curve to be aligned with $\mathcal{T}(\alpha_{n})$. 
\end{enumerate}
We define the cubic B\'{e}zier curve $B_{n}:[0,L_{n}]\rightarrow R^{2}$,  which approximates $C(t)$ in the interval $[t_n,t_{n+1}]$ by 

\begin{multline}
B_{n}(s)  =(1-\frac{s}{L_{n}})^{3}\bar{x}_{n}+ \\
3(1-\frac{s}{L_{n}})^{2}%
\frac{s}{L_{n}}(  \bar{x}_{n}+\lambda_{n}L_{n}\mathcal{T}(\alpha_{n})+\gamma_{n}L_{n}\mathcal{T}(\alpha_{n})^{\perp})  +\\
 3(1-\frac{s}{L_{n}})(  \frac{s}{L_{n}})  ^{2} 
 (  \bar
{x}_{n+1}- \beta_{n}L_{n}\mathcal{T}(\alpha_{n+1})- \\
\delta_{n}L_{n}
\mathcal{T}(\alpha_{n+1})^{\perp})  + 
(  \frac{s}{L_{n}})
^{3}\bar{x}_{n+1},
\end{multline}
where the parameters $\lambda_{n},\gamma_{n},\beta_{n},\delta_{n}\in R$ determine the B\'{e}zier curve control points. $B_{n}(s)$ satisfies: 
\begin{equation} 
\begin{array}{l}
B_{n}(0)=\bar{x}_{n}, \\ 
B_{n}^{\prime}(0)=3(\lambda_{n}\mathcal{T}(\alpha
_{n})+\gamma_{n}\mathcal{T}(\alpha_{n})^{\perp})\\
B_{n-1}(L_{n-1})=\bar{x}_{n},\quad \\ 
B_{n-1}^{\prime}(L_{n-1})=3\left(  \beta
_{n-1}\mathcal{T}(\alpha_{n})+\delta_{n-1}\mathcal{T}(\alpha_{n})^{\perp
}\right).
\end{array}
\end{equation} 

In this formulation of the cubic B\'{e}zier curve, the tangent vector at the end points is expressed in the local reference system given by $\mathcal{T}(\alpha_{n})$ and $\mathcal{T}(\alpha_{n})^{\perp
}$. We assume that  in the regular points ($t_n \notin Corners$), $\gamma_n=\delta_{n-1}=0$ which means that at $\bar x_n$, the tangent vectors to $B_{n-1}(s)$  and to  $B_n(s)$ are aligned with $\mathcal{T}(\alpha_{n})$. In the corner points, the B\'{e}zier curve  parameters are completely free, in particular, in this case, the choice of $\alpha_n$ does not affect the final result because we can obtain any configuration of the B\'{e}zier curve control points independently of the value of $\alpha_n$. 

The active contour model we propose to fit the B\'{e}zier curves to $C(t)$ is given by the  adjustment error functional 
\begin{multline}
    \mathcal{E}(\{t_{n},\alpha_n,\lambda_{n},\gamma_{n},\beta_{n},\delta_{n}\}_{n=1}^{N}):= \\
    \sum_{n=1}^{N} \int_{0}^{L_{n}}\left(  d_{C}(B_{n}(s))+w_{n}\right)  \left\Vert B_{n}%
^{\prime}(s)\right\Vert ds,
\label{eq:energy}
\end{multline}
where $d_C:R^2 \rightarrow R^+$ is the Euclidean distance from any point $\bar x \in R^2$ to $C(t)$ and $w_n \geq 0$. 
By minimizing this functional, we look for a  B\'{e}zier curve collection with minimal distance to $C(t)$ among all B\'{e}zier curves. The weight parameter $w_n \geq 0$ optionally introduces an additional regularization by penalizing the length of $B_n(s)$. In our experiments, we always initially fix $w_n \equiv 0$, which in general provides good results, and optionally increase the value of $w_n$ to remove 
 the undesirable irregularities that can appear in particular  B\'ezier  sections $B_n(s)$ of the curve. Notice that minimization is constrained by the conditions $\gamma_n=\delta_{n-1}=0$ for $t_n \notin Corners$. 
The main advantages of this active contour model are :

\begin{enumerate}
    \item The model enables simultaneous optimization of the location of the interpolation values $t_n$, the orientation of the vectors tangent to the curve in the regular points, and the B\'{e}zier curve parameters. 
    
    \item The explicit introduction, in the B\'{e}zier curve representation,  of the tangent vectors $\mathcal{T}(\alpha_{n})$ at the points $\bar x_n$ allows one to manage in a very simple way the orientation of the tangent vectors at the regular points. Indeed, we only have to fix $\gamma_n=\delta_{n-1}=0$ at the regular points. Moreover, the optimization problem is simplified because we reduce the number of unknowns. 
    \item If we fix the values of $t_n$ and $\alpha_n$, the minimization of (\ref{eq:energy}) to estimate the B\'{e}zier curve parameters can be performed independently in each interval $[t_n,t_{n+1}]$, which strongly simplifies the complexity of the minimization problem. Moreover, it allows for a simple minimization method, where we alternately optimize the B\'{e}zier curve parameters $\alpha_n,\lambda_{n},\gamma_{n},\beta_{n},\delta_{n}$ and the values $t_n$ and $\alpha_n$. 
    \item Given that we always initialize with $w_n=0$, the management of $w_n$ is an optional post-processing of the B\'{e}zier curves. 
\end{enumerate}

\section{Minimization of the adjustment error functional (\ref{eq:energy})}

First, we highlight that to apply our proposed active contour model, we can use any vectorization method to select the initial collection of B\'ezier curves. If the method does not provide information about the set of corner points, we can introduce such information by hand or simply define all interpolation points as corners. 

For $t,t^{\prime}\in\lbrack0,T]$ we denote by
$|C_{t,t^{\prime}}|$ the length of the curve between $t$ and $\ t^{\prime}$
given by
\[
|C_{t,t^{\prime}}|=\int_{t}^{t^{\prime}}\left\Vert C^{\prime}(t)\right\Vert
dt.
\]
Let us denote by $B_{t,t',\alpha,\alpha'}(s)$ the B\'{e}zier curve with end points $C(t)$  and $C(t')$ and with tangent orientation angles $\alpha$ and $\alpha'$ (in particular we have that $B_n(s)=B_{t_n,t_{n+1},\alpha_n,\alpha_{n+1}}(s)$). For fixed values of $t,t',\alpha,\alpha'$, we obtain the B\'{e}zier curve parameters by applying a gradient descent method to the functional: 
\begin{multline}
    E_{t,t',\alpha,\alpha',w}(\lambda,\gamma,\beta,\delta)= \\
    \int_{0}^{|C_{t,t'}|}\! \! \! \! \! \! \! \! \left(  d_{C}(B_{t,t',\alpha,\alpha'}(s))+w\right)\left\Vert B_{t,t',\alpha,\alpha'}%
^{\prime}(s)\right\Vert ds.
\label{eq:e2}
\end{multline}

As initial guess for the active contour model, we use a linear estimate of the B\'{e}zier curve parameters obtained by minimizing the quadratic error
\begin{equation}
Q(\lambda,\gamma,\beta,\delta)=\int_{0}^{|C_{t,t'}|} \! \! \! \! \! \! \! \! \! \! \! \! \left\Vert B_{t,t',\alpha,\alpha'}(|C_{t,s}|)-C(s)\right\Vert ^{2}ds
\label{eq:classic_bezier_estimation}.%
\end{equation}

One can easily show (see, for instance \cite{HHM22}), that the minimization of this quadratic error leads to a linear system of equations. The constraints $\gamma=0$ if $t$ is a regular point or $\delta=0$ if $t'$ is a regular point are directly included in the linear estimation of the parameters and in the gradient descent. To compute the distance function $d_C(\bar x)$ we use an efficient algorithm presented in \cite{HHA21}. 

To minimize (\ref{eq:energy}) with respect to $\{t_n\}$ and $\{\alpha_n\}$, first we take into account that if $t_{n-1}$, $t_{n+1}$, $\alpha_{n-1}$ and $\alpha_{n+1}$ are fixed, we can find the optimal values for $t_n$ and $\alpha_n$ by minimizing the functional : 
\begin{multline}
    E_n(t,\alpha)=E_{t_{n-1},t,\alpha_{n-1},\alpha,w_{n-1}}(\lambda,\gamma,\beta,\delta) \ + \\
    E_{t,t_{n+1},\alpha,\alpha_{n+1},w_{n}}(\lambda,\gamma,\beta,\delta),
\label{eq:e3}
\end{multline}
where for each choice of $t$ and $\alpha$, the values of $\lambda,\gamma,\beta,\delta$ are, in each case, the one obtained minimizing (\ref{eq:e2}). Therefore, the minimization of (\ref{eq:energy}) can be obtained by iterations based on the following steps: 

\begin{enumerate}
    \item For each $n=1,..,N$:  
      
          we update $t_n=t_{min}$ and $\alpha_n=\alpha_{min}$  where $(t_{min},\alpha_{min}) = \text{arg min } E_n(t,\alpha)$ for $(t,\alpha) \in [t_n-2,t_n+2]\times[\alpha_n-r_{\alpha},\alpha_n+r_{\alpha}]$ where $r_{\alpha}$ is fixed to 4 degrees (in the case of corner points we do not minimize with respect to $\alpha$). 
       
    \item If after updating $\{t_n\}$ and $\{\alpha_n\}$ the functional (\ref{eq:energy}) is
significantly reduced, we repeat the previous step to update these values again; otherwise, we stop the iteration. 
\end{enumerate}

\section{Experiments}

 For comparison purposes, in addition to the standard vectorization algorithms provided by Inkscape and Adobe Illustrator we also use a basic vectorization algorithm that we name "curvature method", presented in the Appendix, based on the curvature of the silhouette contour. This basic method allows us to compare the usual linear method (\ref{eq:classic_bezier_estimation}) to estimate the parameters of the B\'ezier curves and its improvement using the proposed active contour model.

For simplicity, we make experiments with silhouette contours given by a single Jordan curve $C:[0,T]\rightarrow R^{2}$. In the case of multiple Jordan curves, we would apply the active contour model independently to each Jordan curve. We denote by $\mathcal{B}=\{B_{n}(s):[0,L_{n}]\rightarrow R^{2}\}$ the approximation of $C$ by a collection
of B\'{e}zier curves. To measure the quality of this approximation, we use the average distance between $\mathcal{B}$ and $C$ given by
\begin{equation}
d(\mathcal{B},C)=\frac{\sum_{n}\int_{0}^{L_{n}}d_{C%
}(B_{n}(s))\left\Vert \frac{d}{ds}B_{n}(s)\right\Vert ds}{
\sum_{n}\int_{0}^{L_{n}}\left\Vert \frac{d}{ds}B_{n}(s)\right\Vert ds},%
\label{eq:dBC}
\end{equation}
and the average distance between $C$ and $\mathcal{B}$ given by: 
\begin{equation}
d(C,\mathcal{B})=\frac{\int_{0}^{T}d_{\mathcal{B}
}(C(t))\left\Vert \frac{d}{dt}C(t)\right\Vert dt}{\int_{0}^{T
}\left\Vert \frac{d}{dt}C(t)\right\Vert dt}.
\label{eq:dCB}
\end{equation}

It is very important to check the values of both distances, because it could happen that the B\'{e}zier curves  $\mathcal{B}$ are close to $C$ ($d(\mathcal{B},C)$ is small)  but that $\mathcal{B}$ does not adequately cover $C$ (that is, $d(C,\mathcal{B})$  is large). The distance $d(\mathcal{B},C)$ is closely related to the proposed active contour model because our method aims at optimizing the B\'{e}zier curves by minimizing the term in the numerator of (\ref{eq:dBC}). Notice that it would be possible to use a more symmetric active contour formulation using the combined energy: 
\begin{multline}
    E(\mathcal{B})=\int_{0}^{T}d_{\mathcal{B}}(C(t))\left\Vert \frac{d}{dt}C(t)\right\Vert dt \ + \\\sum_{n}\int_{0}^{L_{n}}d_{C}(B_{n}(s))\left\Vert \frac{d}{ds}B_{n}(s)\right\Vert ds.
\end{multline}

The main technical limitation of this approach is that each evaluation of $E(\mathcal{B})$ requires the computation of the distance function $d_{\mathcal{B}}(\bar x)$. Given that  minimization of $E(\mathcal{B})$ requires a very large number of evaluations of $E(\mathcal{B})$ for different choices of $\mathcal{B}$, the computational cost of the algorithm would be very high. However, as we show in the experiments, with the proposed active contour model, the values obtained for $d(\mathcal{B},C)$ and $d(C,\mathcal{B})$ are reasonably similar.

We denote by $\mathcal{B}_{\kappa}^0$ the initial B\'ezier curves obtained by the basic curvature based vectorization algorithm presented in the appendix using the linear method (\ref{eq:classic_bezier_estimation}) to compute the B\'ezier curves parameters, and by $\mathcal{B}_{\kappa}^{\infty}$ the B\'ezier curves obtained as the asymptotic state of the active contour model using as initial guess $\mathcal{B}_{\kappa}^0$, by comparing $d(\mathcal{B}_{\kappa}^0,C)$ with  $d(\mathcal{B}_{\kappa}^{\infty},C)$ or $d(C,\mathcal{B}_{\kappa}^0)$ with  $d(C,\mathcal{B}_{\kappa}^{\infty})$ we measure the ability of the active contour model to improve the quality of the approximation of the silhouette contour by the B\'ezier curves obtained using the usual linear method (\ref{eq:classic_bezier_estimation}) to compute the B\'ezier curve parameters.

Inkscape is a powerful open source vector graphics editor widely used for creating and editing vector illustrations, logos, icons, diagrams, and complex designs. We use the Inkscape vectorization algorithm (see \cite{Selinger2003PotraceA}) to provide the initial guess of our active contour model. In all experiments, we use the Inkscape interface to obtain the silhouette vectorization using the maximum optimization value allowed, then simplify the collection of B\'{e}zier curves; we export the result as an SVG file and we read this file to get the initial B\'{e}zier curves that we name $\mathcal{B}_{I}^0$. We denote by $\mathcal{B}_{I}^{\infty}$ the asymptotic state of the active contour model using as an initial guess the original Inkscape vectorization $\mathcal{B}_{I}^0$. 

Adobe Illustrator is a widely used professional graphic editor. We used the Adobe Illustrator image trace tool to vectorize the image silhouettes we use in our experiments, and we used the Adobe Illustrator simplification tool to reduce the number of end points of the B\'ezier curves. We denote by $\mathcal{B}_{A}^{\infty}$ the asymptotic state of the active contour model using as initial guess $\mathcal{B}_{A}^0$ (the original Adobe Illustrator vectorization).   

In the Inkscape and Adobe Illustrator SVG file outputs, we have no information about corner/regular points, so we assume that all points have two half-tangents and, therefore, are "corners".

Our main goal in the experiments is to check if the active contour model is able to significantly reduce the values of $d(\mathcal{B}^0,C)$ and $d(C,\mathcal{B}^0)$. In table \ref{tab:table}, we present the results obtained for three silhouettes. Each silhouette is given by a 1024$\times$1024 image. In all experiments, we observe a significant reduction in the distances between $C(t)$ and the B\'ezier curves before and after applying the proposed active contour model. Notice that the number of nodes (end points) of the B\'ezier curves can be different among the vectorization methods but we emphasize that for our purpose it is not relevant because we are interested in the ability of the active contour models to reduce the distance between the silhouette contour and the B\'ezier curves for a given vectorization method.

\begin{table*}[]
    \centering
\begin{tabular}
[c]{|l|l|l|l|l|l|l|l|}\hline
{\footnotesize	 Silhouette} & {\footnotesize	Nodes} & {\footnotesize	$d(\mathcal{B}_{\kappa}^0,C)$ }& {\footnotesize	$d(\mathcal{B}_{\kappa}^{\infty}
,C)$} & {\footnotesize	 Var. Perc.} & {\footnotesize	$d(C,\mathcal{B}_{\kappa}^0)$ }& {\footnotesize	$d(C,\mathcal{B}_{\kappa}^{\infty})$} & {\footnotesize	Var. Perc.}\\\hline
 \includegraphics[height=0.05\linewidth]{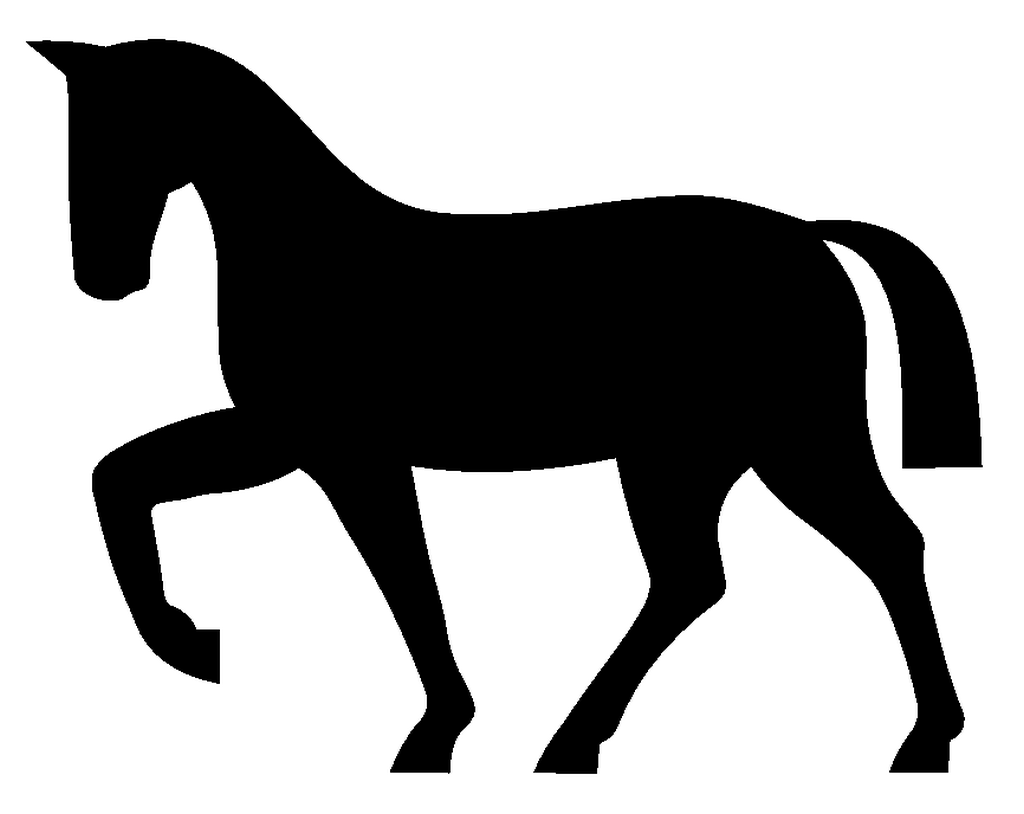}  & 46 & 1.08 & 0.57 & -47.33\% & 1.02 & 0.61 & -40.47\% \\\hline
 \includegraphics[height=0.05\linewidth]{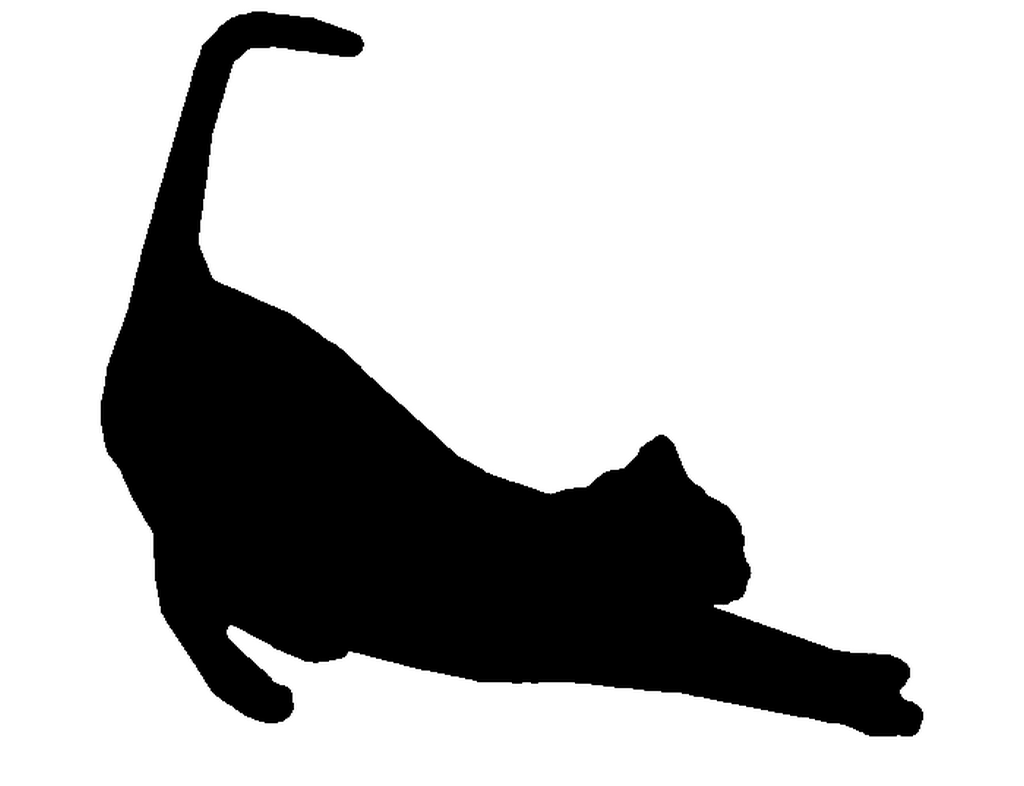}    & 28 & 1.48 & 0.68 & -54.10\% & 1.46 & 0.68 & -53.24\% \\\hline
\includegraphics[height=0.05\linewidth]{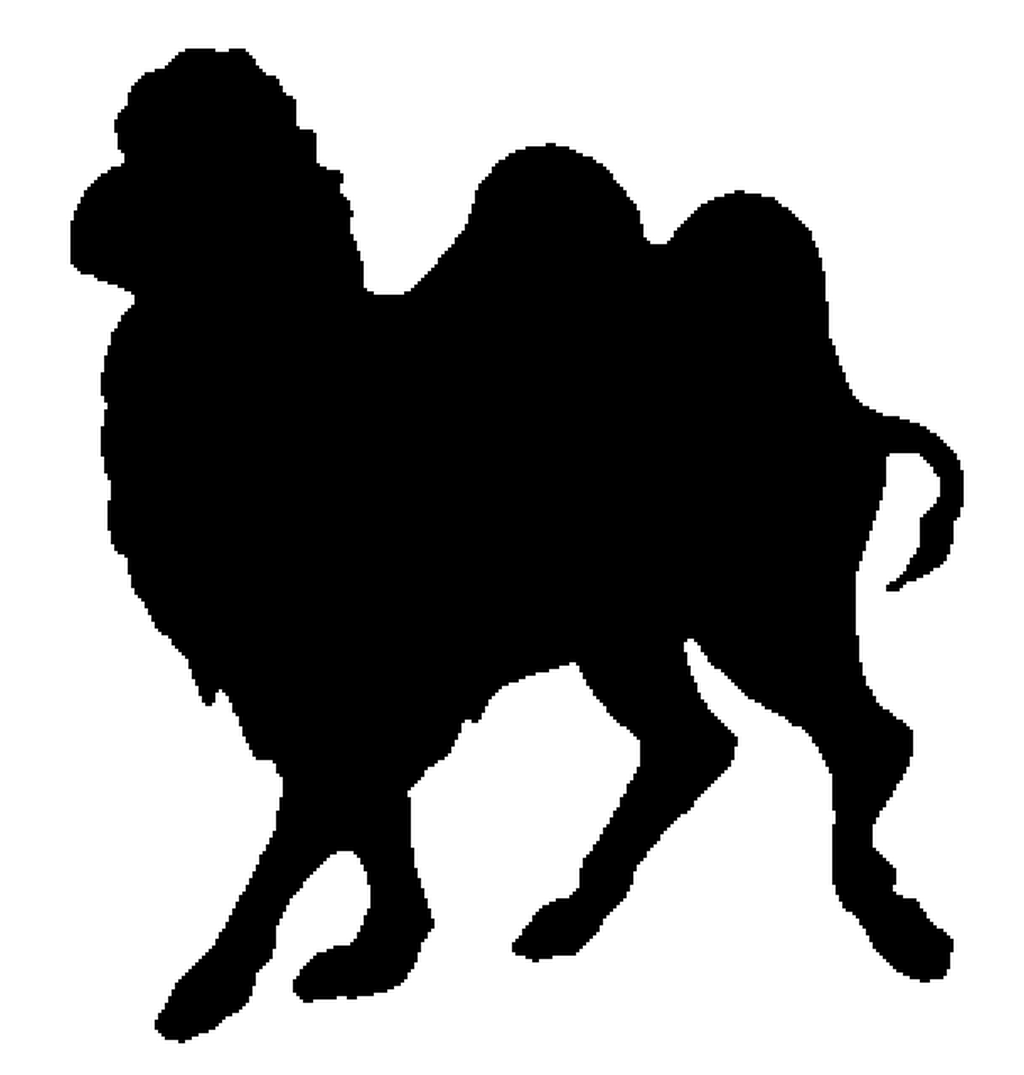}   & 61 & 1.39 & 0.86 & -37.71\% & 1.62 & 1.12 & -30.75\% \\\hline
\end{tabular} \medskip

\begin{tabular}
[c]{|l|l|l|l|l|l|l|l|}\hline
{\footnotesize	Silhouette} & {\footnotesize	Nodes} & {\footnotesize	$d(\mathcal{B}_{I}^0,C)$ } & {\footnotesize	$d(\mathcal{B}_{I}^{\infty}
,C)$ }&  {\footnotesize	Var. Perc. }& {\footnotesize	$d(C,\mathcal{B}_{I}^0)$ }& {\footnotesize	$d(C,\mathcal{B}_{I}^{\infty})$} & {\footnotesize	Var. Perc.}\\\hline
 \includegraphics[height=0.05\linewidth]{caballo.png} & 48 & 0.97 & 0.82 & -15.46\% & 0.86 & 0.80 & -7.11\% \\\hline
 \includegraphics[height=0.05\linewidth]{cat.png}   & 32 & 0.98 & 0.59 & -39.50\% & 1.09 & 0.59 & -45.58\% \\\hline
\includegraphics[height=0.05\linewidth]{camel.png}  & 111 & 0.92 & 0.61 & -34.24\% & 1.15 & 0.77 & -33.17\% \\\hline
\end{tabular}
\medskip

\begin{tabular}
[c]{|l|l|l|l|l|l|l|l|}\hline
{\footnotesize	Silhouette} & {\footnotesize	 Nodes} & {\footnotesize	 $d(\mathcal{B}_{A}^0,C)$} & {\footnotesize	 $d(\mathcal{B}_{A}^{\infty}
,C)$} & {\footnotesize	  Var. Perc.} & {\footnotesize	 $d(C,\mathcal{B}_{A}^0)$} & {\footnotesize	 $d(C,\mathcal{B}_{A}^{\infty})$} & {\footnotesize	 Var. Perc.}\\\hline
 \includegraphics[height=0.05\linewidth]{caballo.png}  & 53 & 0.97 & 0.62 & -36.76\% & 0.83 & 0.60 & -27.47\% \\\hline
 \includegraphics[height=0.05\linewidth]{cat.png}  & 25 & 1.89 & 1.08 & -42.85\% & 1.70 & 1.00 & -40.90\% \\\hline
\includegraphics[height=0.05\linewidth]{camel.png}  & 117 & 1.15 & 0.72 & -37.76\% & 1.18 & 0.88 & -25.66\% \\\hline
\end{tabular}

    \caption{ From top to bottom we show the results obtained for the basic curvature method given by $\mathcal{B}_{\kappa}^0$ (initial guess for the active contour model) and $\mathcal{B}_{\kappa}^{\infty}$ (final result), the Inkscape software, $\mathcal{B}_{I}^0$ and $\mathcal{B}_{I}^{\infty}$ and the Adobe Illustrator software  $\mathcal{B}_{A}^0$ and $\mathcal{B}_{A}^{\infty}$. In each row, from left to right, we show: a snapshot of the silhouette used in the experiment, the number of nodes (end points) of the B\'ezier curves, the values of the distances between the curve $C$ and the corresponding B\'ezier curves and the variance percentage of the distance before and after the application of the active contour model. 
    } 
    \label{tab:table}
\end{table*}

\begin{figure*}
    \begin{center}
    \begin{tabular}{|c|c|}\hline
        Horse Silhouette & $\mathcal{B}_{\kappa}^{\infty}$ (curvature method)\\ 
        \includegraphics[trim= 0 0 0 0 ,clip, width=0.25\linewidth]{caballo.png} &
        \includegraphics[trim= 20 20 20 20 ,clip, width=0.25\linewidth]{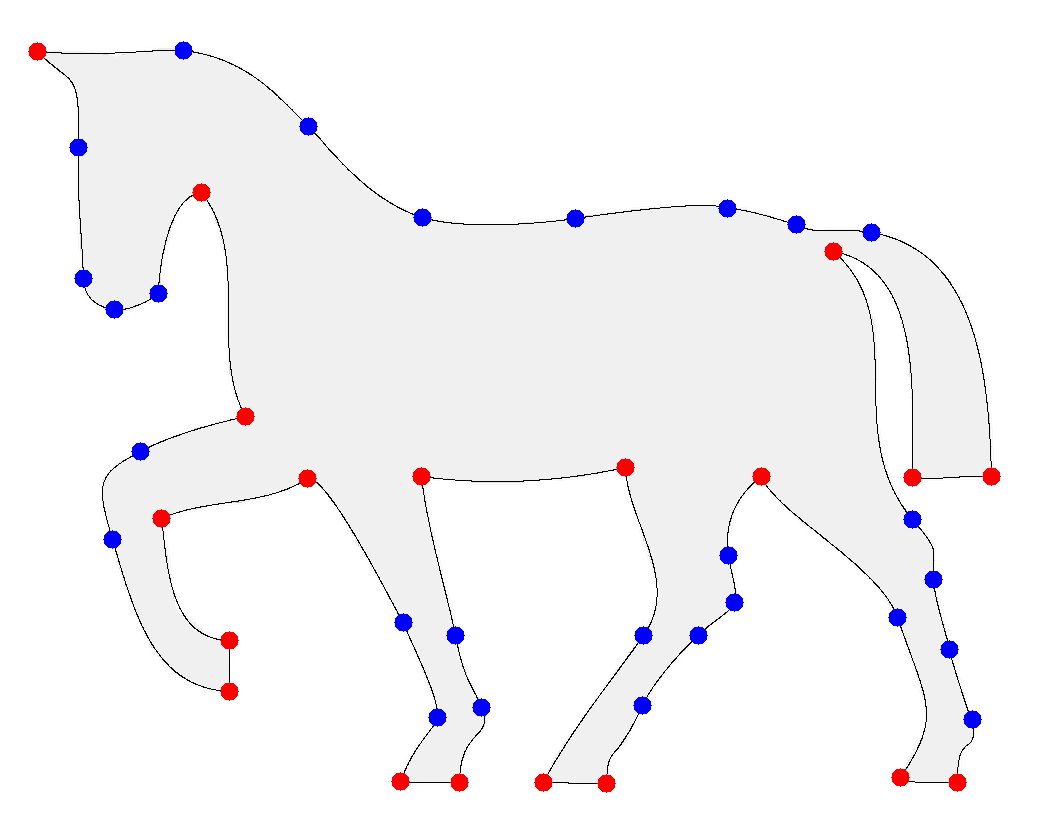}  \\ \hline
        $\mathcal{B}_{I}^{\infty}$ (Inkscape) & $\mathcal{B}_{A}^{\infty}$ (Adobe Illustrator)\\ 
        \includegraphics[trim= 20 20 20 20 ,clip, width=0.25\linewidth]{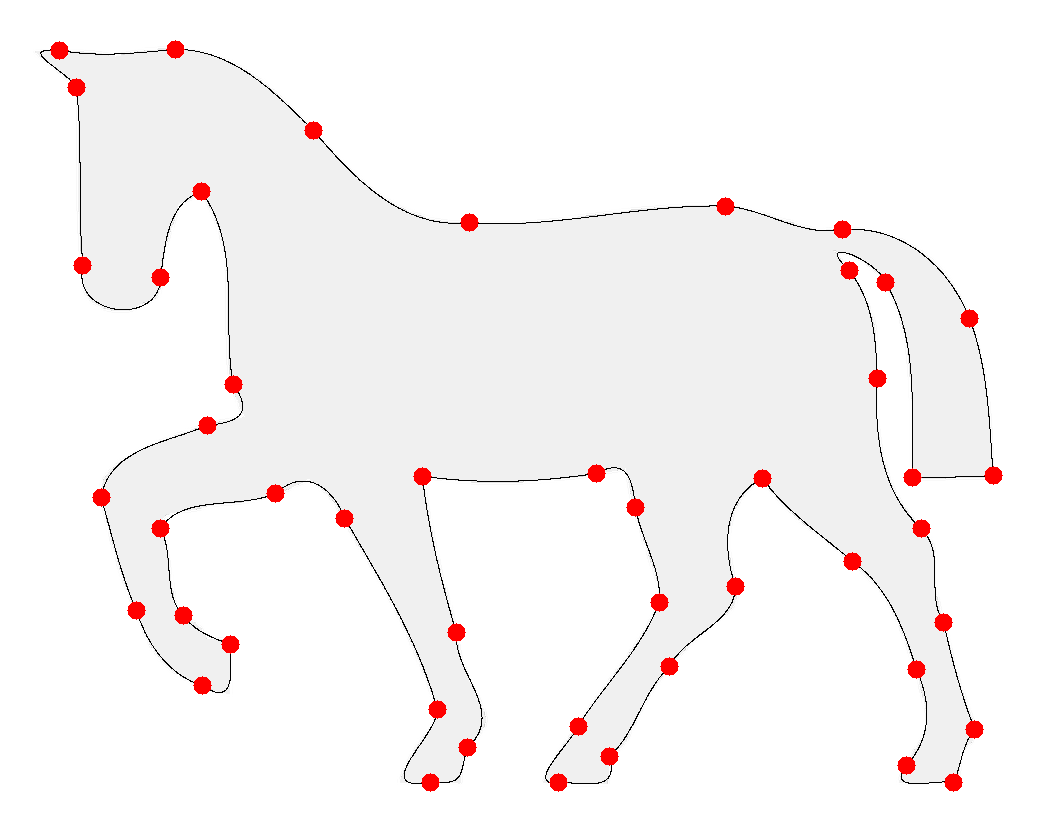} &
        \includegraphics[trim= 20 20 20 20 ,clip, width=0.25\linewidth]{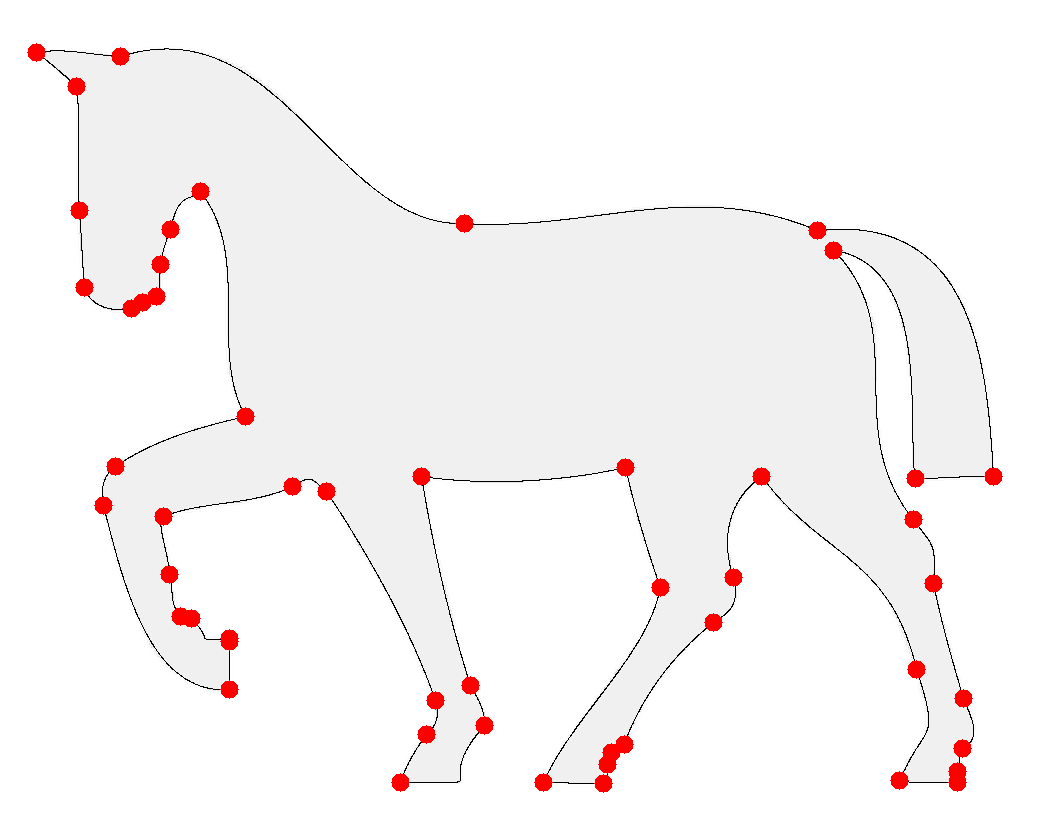} \\ \hline
    \end{tabular}
    \caption{ \label{fig:horse} Horse silhouette and the asymptotic state of the active contour models using as initial guess the vectorization provided by different methods. Blue circles represent regular points and red circles corners.}
    \end{center}
\end{figure*}

\begin{figure*}
    \begin{center}
    \begin{tabular}{|c|c|}\hline
        Cat Silhouette & $\mathcal{B}_{\kappa}^{\infty}$ (curvature method)\\ 
        \includegraphics[trim= 0 0 0 0 ,clip, width=0.25\linewidth]{cat.png} &
        \includegraphics[trim= 20 20 20 20 ,clip, width=0.25\linewidth]{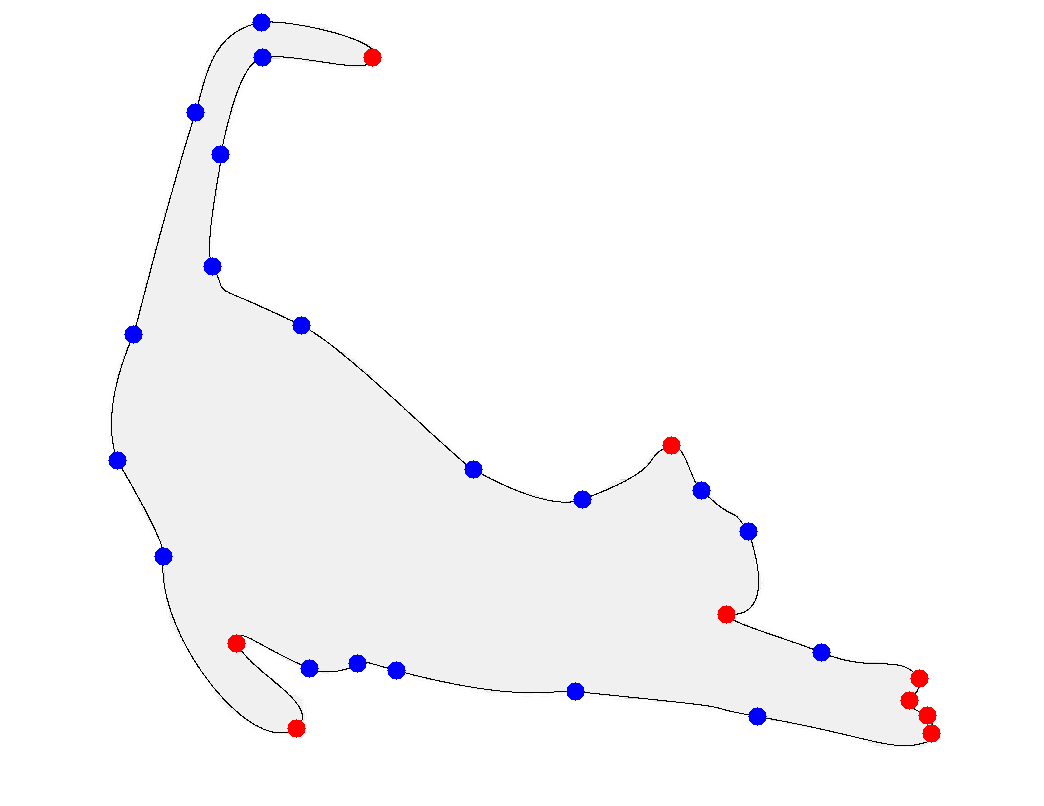}  \\ \hline
        $\mathcal{B}_{I}^{\infty}$ (Inkscape) & $\mathcal{B}_{A}^{\infty}$ (Adobe Illustrator)\\ 
        \includegraphics[trim= 20 20 20 20 ,clip, width=0.25\linewidth]{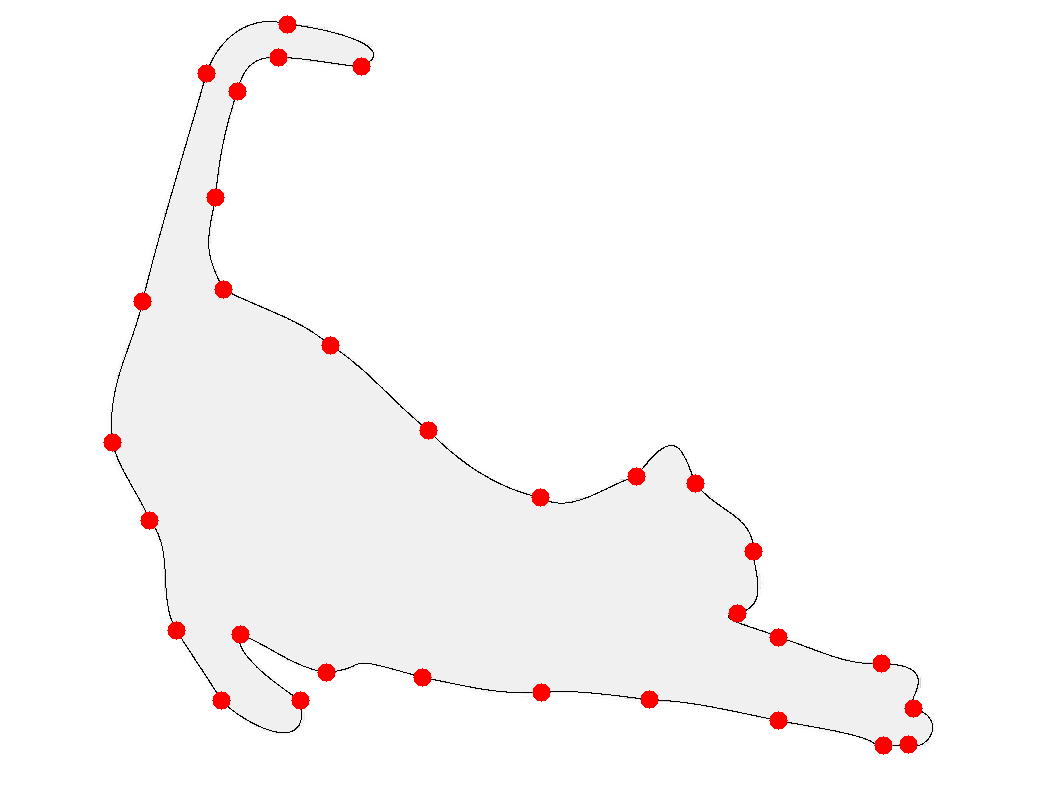} &
        \includegraphics[trim= 20 20 20 20 ,clip, width=0.25\linewidth]{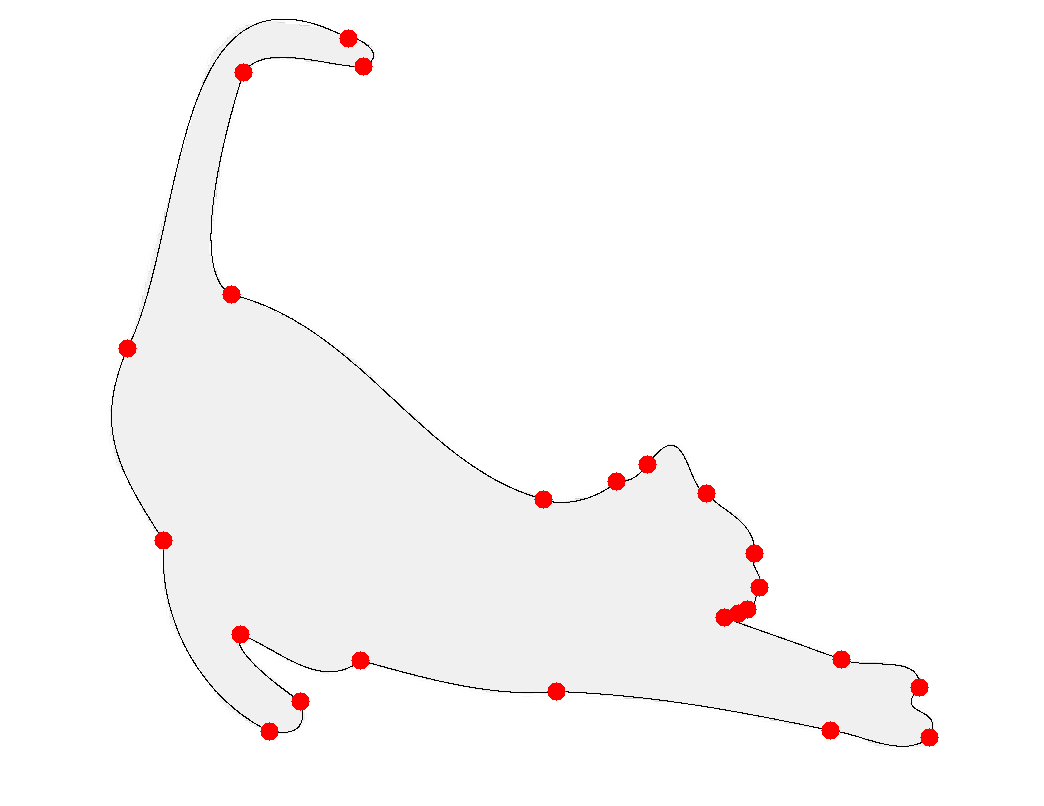} \\ \hline
    \end{tabular}
    \caption{ \label{fig:cat} Cat silhouette and the asymptotic state of the active contour models using as initial guess the vectorization provided by different methods. Blue circles represent regular points and red circles corners.}
    \end{center}
\end{figure*}

\begin{figure*}
    \begin{center}
    \begin{tabular}{|c|c|}\hline
        Camel Silhouette & $\mathcal{B}_{\kappa}^{\infty}$ (curvature method)\\ 
        \includegraphics[trim= 0 0 0 0 ,clip, width=0.25\linewidth]{camel.png} &
        \includegraphics[trim= 20 20 20 20 ,clip, width=0.25\linewidth]{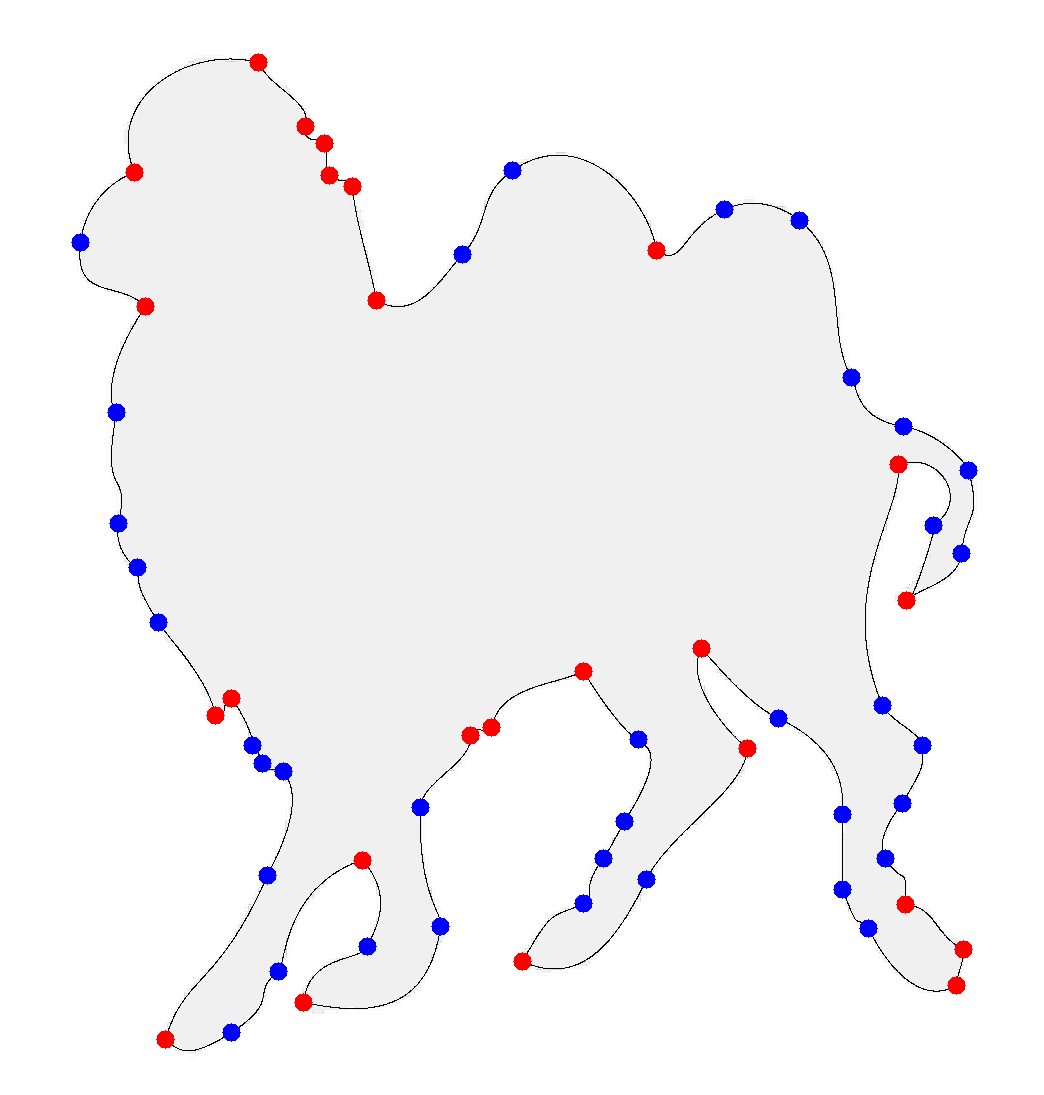}  \\ \hline
        $\mathcal{B}_{I}^{\infty}$ (Inkscape) & $\mathcal{B}_{A}^{\infty}$ (Adobe Illustrator)\\ 
        \includegraphics[trim= 20 20 20 20 ,clip, width=0.25\linewidth]{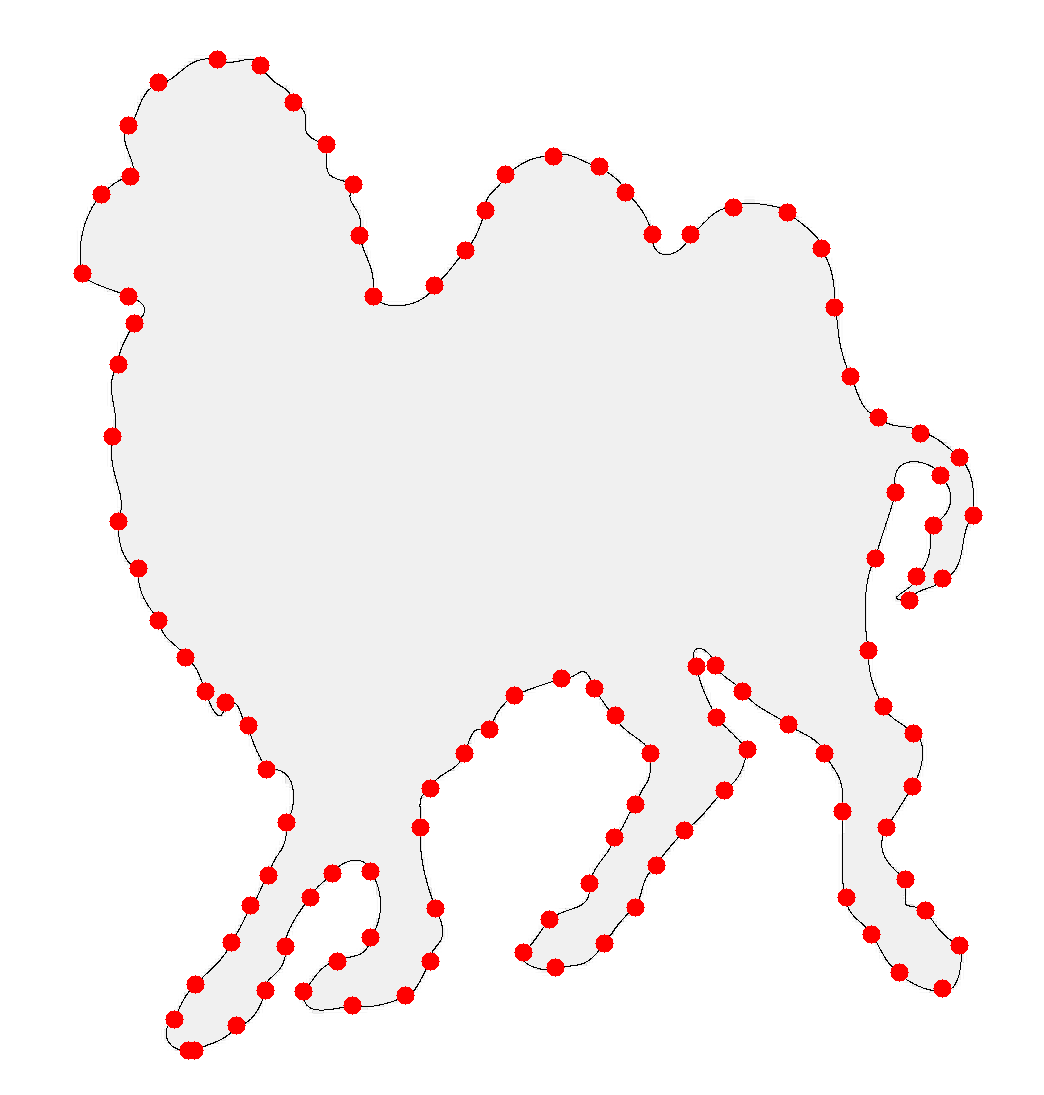} &
        \includegraphics[trim= 20 20 20 20 ,clip, width=0.25\linewidth]{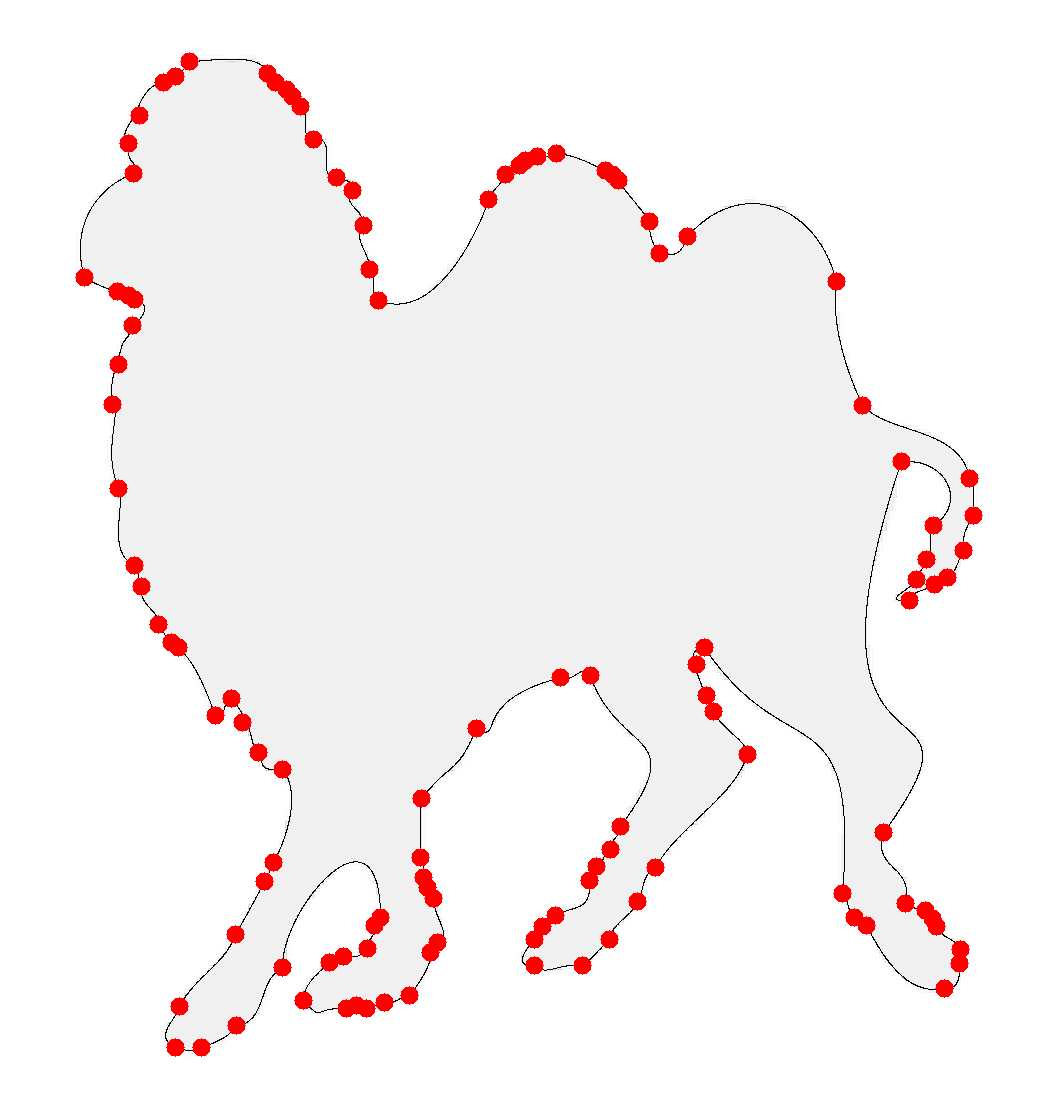} \\ \hline
    \end{tabular}
    \caption{ \label{fig:camel} Camel silhouette and the asymptotic state of the active contour models using  the vectorization provided by different methods as initial guess. Blue circles represent regular points and red circles corners.}
    \end{center}
\end{figure*}

In Figs. \ref{fig:horse}, \ref{fig:cat} and \ref{fig:camel} we show the silhouettes we use in our experiments and the asymptotic state of the active contour model for the different vectorization methods. We highlight that the collection of B\'ezier curves end-points is very different for the different methods due to the different strategies to select such  end-points.

\begin{figure*}
    \begin{center}
    \begin{tabular}{|c|c|c|}\hline
        curvature method  & 
        curvature method &
        curvature method \\ 
        Horse zoom & 
        Horse zoom  &
        Horse zoom  \\ 
        \includegraphics[width=0.20\linewidth]{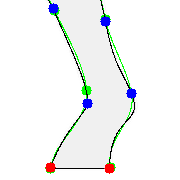} &
         \includegraphics[width=0.20\linewidth]{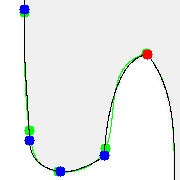} &
          \includegraphics[width=0.20\linewidth]{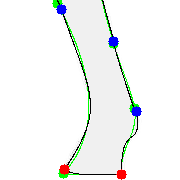} \\ \hline
        Inkscape  & Inkscape  & Inkscape  \\
        Cat zoom & Cat zoom & Cat zoom \\
       \includegraphics[width=0.20\linewidth]{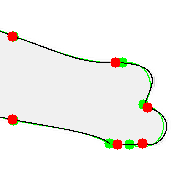} &
         \includegraphics[width=0.20\linewidth]{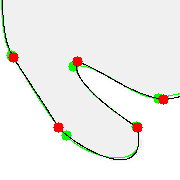} &
          \includegraphics[width=0.20\linewidth]{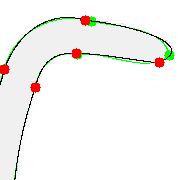} \\ \hline
        Adobe Illustrator  & Adobe Illustrator  & Adobe Illustrator  \\
        Camel zoom & Camel zoom & Camel zoom \\
       \includegraphics[width=0.20\linewidth]{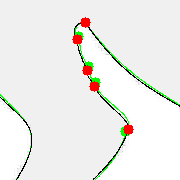} &
         \includegraphics[width=0.20\linewidth]{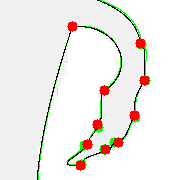} &
          \includegraphics[width=0.20\linewidth]{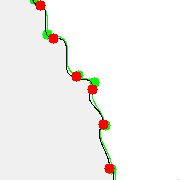} \\ \hline
    \end{tabular}
    \caption{ \label{fig:zoom} We present a zoom of different parts of the silhouettes. The asymptotic state of the active contour model is represented by blue circles (regular points) or red circles (corners) with the drawing of the B\'ezier curves in black. The original silhouette vectorization is represented in green. }
    \end{center}
\end{figure*}

In Fig. \ref{fig:zoom}, we present some details on the comparison of the initial vectorization  with the one provided by the active contour model.  We observe that the active contour model provides a better matching of the B\'ezier curves and silhouette contour and that the original position of the B\`ezier curves end points can be shifted by the active contour model. 

\begin{figure*}
    \centering
    \includegraphics[trim=80 20 420 570,clip, width=0.3\linewidth]{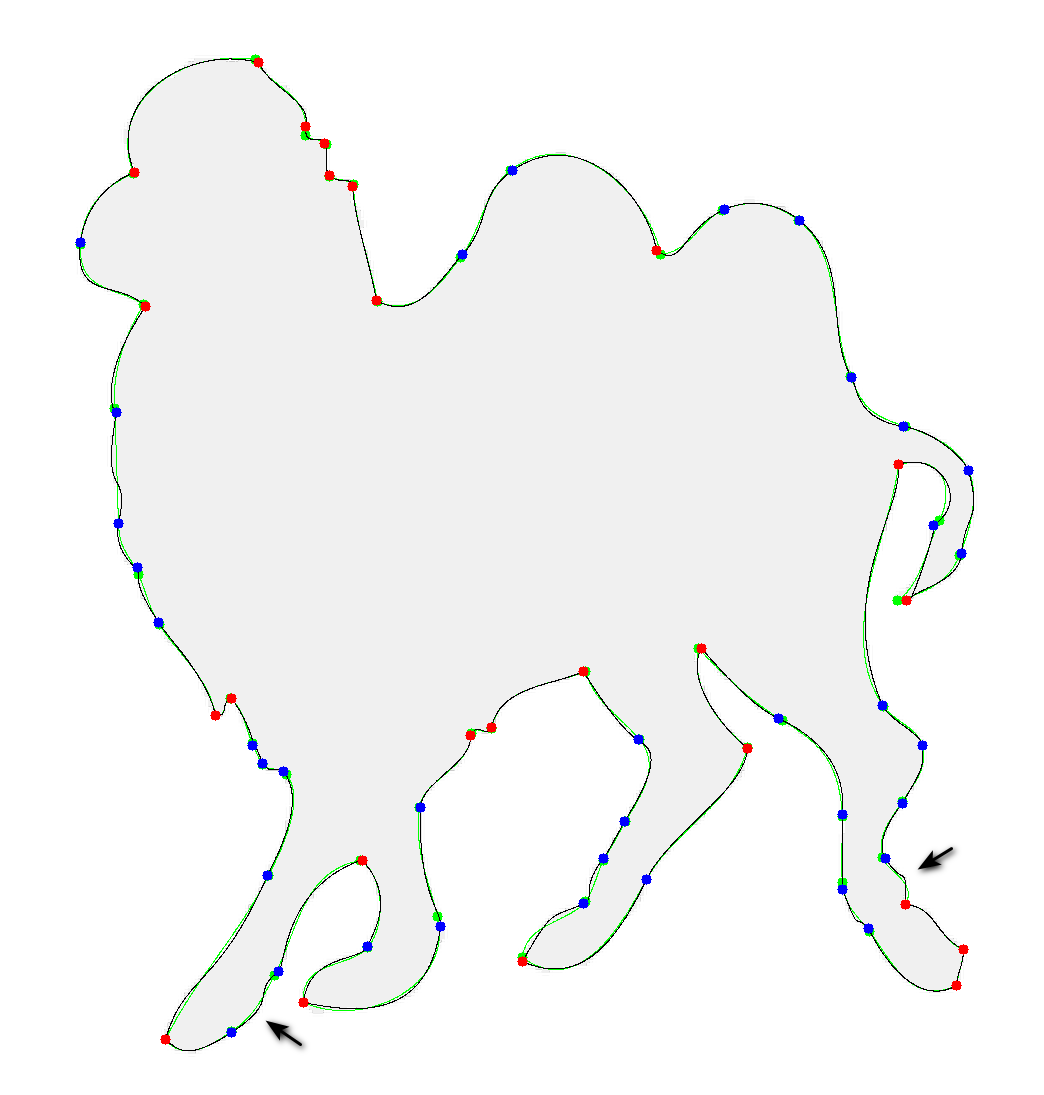}
    \includegraphics[trim=130 20 700 950,clip, width=0.3\linewidth]{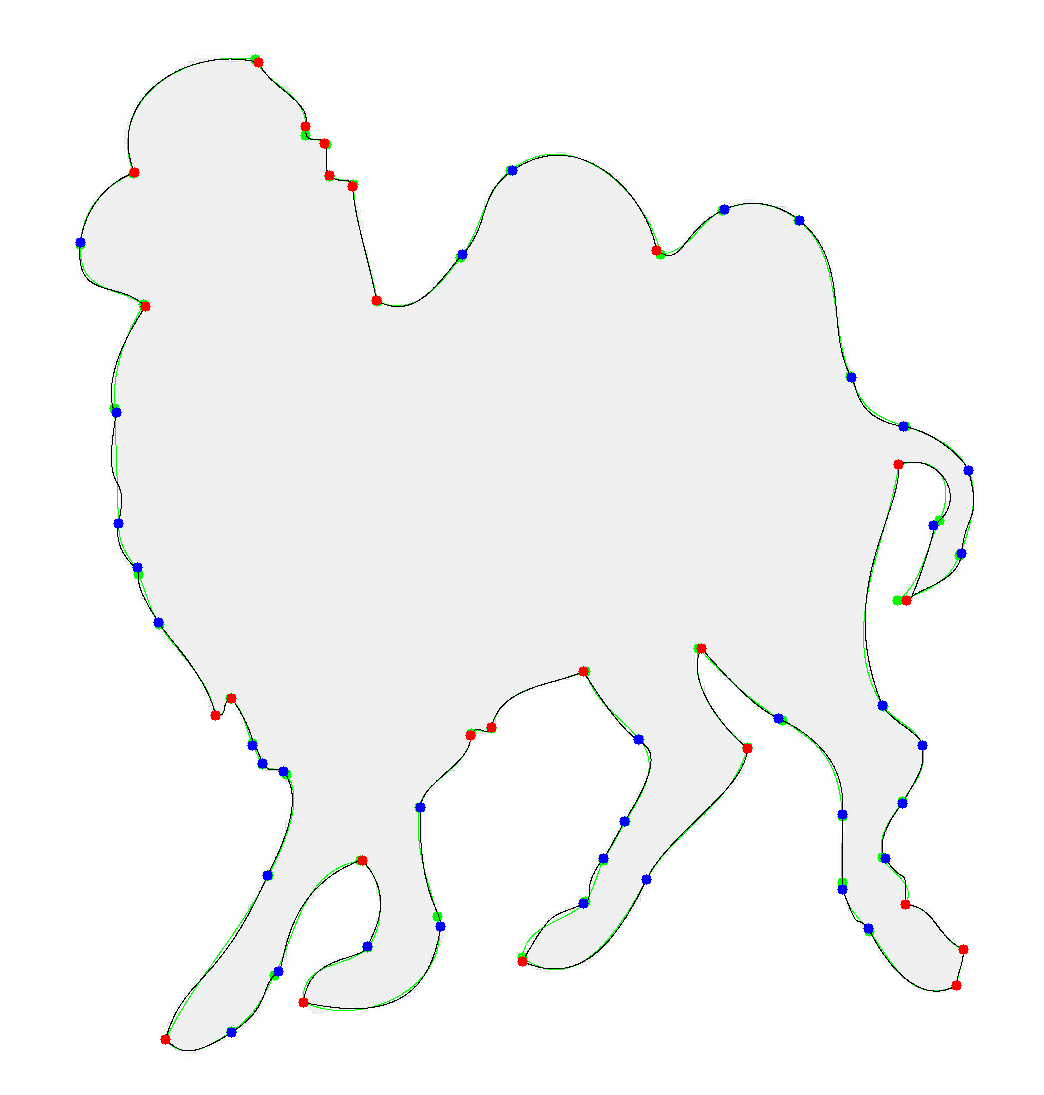} \\
    \includegraphics[trim=500 100 40 500,clip, width=0.25\linewidth]{camel_bezier1original.png}
    \includegraphics[trim=800 170 40 830,clip, width=0.35\linewidth]{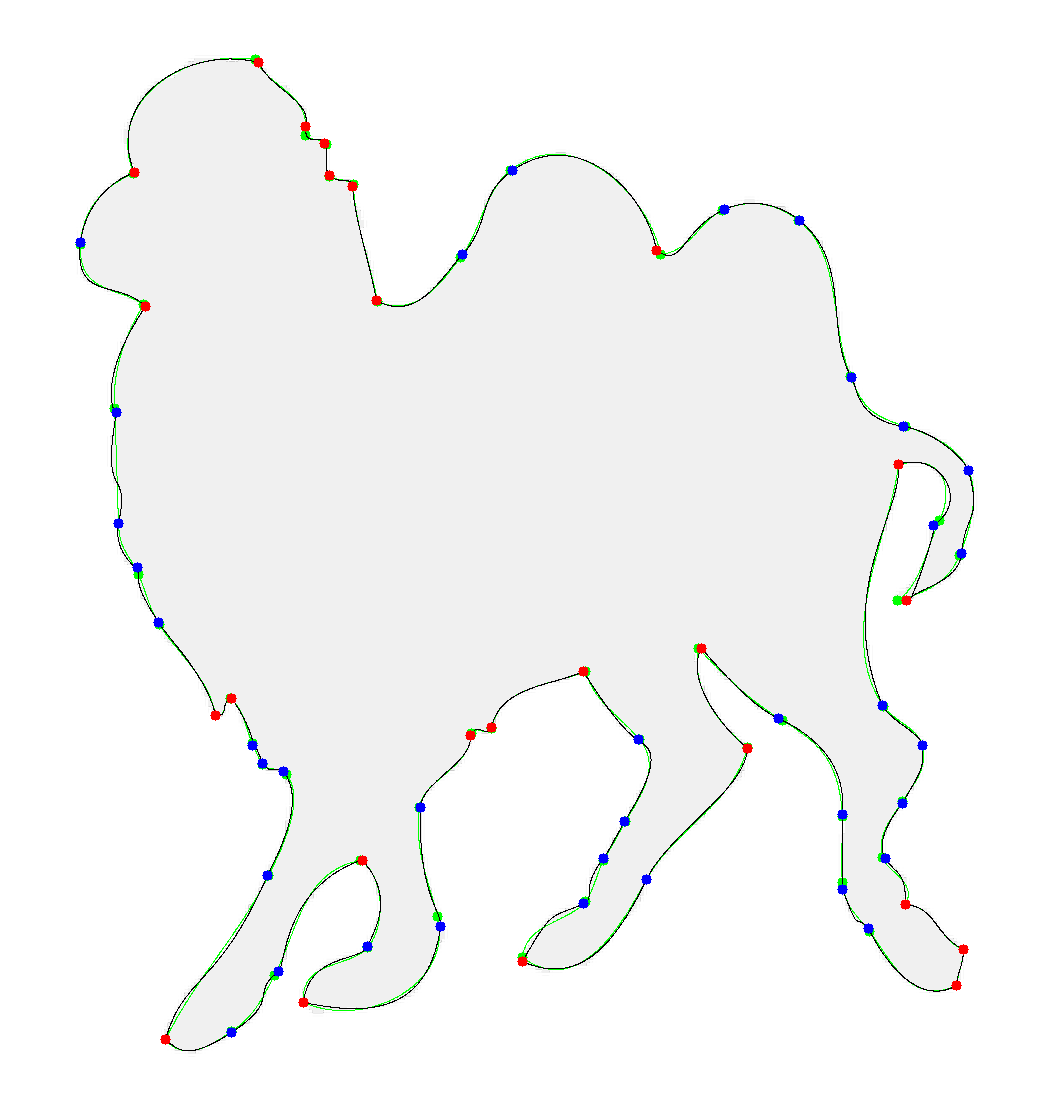}
    \caption{ On the left, we present a zoom of the vectorization of the camel silhouette using the basic curvature method with the regularization default value $w_n \equiv 0$. On the right, we present the result obtained using $w_n=30$ as regularization parameter for the marked B\'ezier curve section. 
    \label{fig:regularization}
    }
    
\end{figure*}

In Fig. \ref{fig:regularization}, we illustrate the effect of the regularization parameter $w_n$ (by default $w_n \equiv 0$). We show that when $w_n>0$ in a B\'ezier curve section, the energy minimization of the active contour model produces a regularization of the B\'ezier curve by minimizing its length. 

\clearpage

\section{Conclusion}
Based on the geodesic active contour formulation introduced by Caselles et al. in \cite{CKS97}, we propose an active contour model that optimizes the approximation of a curve $C(t)$ by a collection of cubic B\'{e}zier curves $\{B_n(s)\}$. By minimizing the associated energy, this model is able to optimize the location of points on $C(t)$ used as end points of the B\'{e}zier curves, the orientation of the tangent vectors in the regular points ($t_n \notin Corners$) and the parameters of the B\'{e}zier curves. We tested our method using as an initial guess the vectorization provided by a basic vectorization method based on the silhouette contour curvature and the world-class graphics software Inkscape and Adobe Illustrator. In all cases, using the proposed method, we observed a significant reduction in the distance between $C(t)$ and $\{B_n(s)\}$. 
The additional regularity parameters $w_n$ included in the active contour model allow for an additional regularization of each B\'{e}zier curve section (if needed).

\bibliography{citations.bib}

\section*{Appendix. A basic vectorization algorithm based on the curvature of the silhouette contour}

We denote by $\mathcal{N}_{r}(t)$, a neighborhood in $C(t)$ of radius $r\geq0$ defined by
\begin{multline*}
\mathcal{N}_{r}(t):= \\ 
\{t^{\prime}\in\lbrack0,T]:|C_{t,t^{\prime}}|\leq
r\}\cup\{t^{\prime}\in\lbrack0,T]:|C_{t^{\prime},t}|\leq r\}.
\end{multline*}
Given a scale parameter $\sigma>0$  we define a
cornerness measure $\kappa_{\sigma}(C(t))$ by
\begin{multline*}
\kappa_{\sigma}(C(t)):= \\
\frac{\left(  C(t+h_{\sigma})-C(t),C(t-h_{\sigma
}^{\prime})-C(t)\right)  }{\left\Vert C(t+h_{\sigma})-C(t)\right\Vert
\left\Vert C(t-h_{\sigma}^{\prime})-C(t)\right\Vert }+1.
\end{multline*}
where $(.,.)$ represents the usual dot product and $h_{\sigma},h_{\sigma
}^{\prime}\geq0$ satisfy $|C_{t,t+h_{\sigma}}|=|C_{t,t-h_{\sigma}^{\prime}%
}|=\sigma.$ This definition is explained by the remark that if $C(t)$ has a perfect corner at $t$ of angle
$\alpha,$ then $\kappa_{\sigma}(C(t))=\cos(\alpha)+1.$ The parameter $\sigma$
is also used to compute the initial tangent vector
$\mathcal{T}(\alpha)$: we fix $\alpha$ as the orientation angle of $C_{\sigma}^{\prime}(t)$, where  $C_{\sigma}(t)$ is  a Gaussian convolution of $C(t)$ with standard deviation $\sigma.$

The algorithm starts by iteratively defining the set $Corners$. At each iteration we compute 

\begin{equation}
t^{\prime}:=\underset{t\in\lbrack0,T]-\cup_{t\in Corners}\mathcal{N}_{r}%
(t)}{\arg\max}\kappa_{\sigma}(C(t))%
\end{equation}
if $\kappa_{\sigma}(C(t^{\prime})) \geq \kappa_{\min}$, then we include $t^{\prime}$ in $Corners$ and we iterate.   The threshold $\kappa_{\min}$ is a fixed method parameter and $r=MinLength$. In short, $Corners$ is a set of points with large enough cornerness $\kappa_{\sigma}(C(t))$ and separated from each other by a distance larger than $MinLength$, where  $MinLength$ is a parameter of the algorithm representing the minimun length allowed for $|C_{t_n,t_{n+1}}|$. 
Next, we iteratively introduce additional regular points. At each iteration, we first estimate $B_n(s)$ using the linear estimation obtained by minimizing  (\ref{eq:classic_bezier_estimation}) with the current selected points and we compute
\begin{equation}
{\bar x}^{\prime}:={\arg\max}_{s, 1\leq n \leq N}d_C(B_n(s))
\end{equation}
then, if $d_C({\bar x}^{\prime})>MaxDist$ we add, as regular point, the value 
\begin{equation}
t^{\prime}=\underset{t\in\lbrack0,T]-\cup_{1\leq n \leq N}\mathcal{N}_{2r}%
(t_{n})}{\arg\min}\Vert C(t) -  {\bar x}^{\prime} \Vert)%
\end{equation}
and we iterate again. That is, we include regular points in locations where the B\'{e}zier curves is far from $C(t)$.

\noindent The main parameters of the algorithm are : 

\begin{enumerate}
\item $MaxDist$ : the maximum distance allowed between the B\'{e}zier curves and the original curve $C(t).$ This is the most important parameter of the algorithm and determines the number of end points of the B\'ezier curves. In all the experiments presented, we fix $MaxDist=6$ which provides a few interpolation values allowing a better illustration of the performance of the proposed active contour model. 

\item $MinLength$ : the minimum length distance (along the curve $C(t)$) between the selected points.  In all the experiments presented, we fix $MinLength=25$.

\item $\sigma$ : scale parameter used to estimate the cornerness measure and the tangent vector to $C(t)$. In all the experiments presented, we fix $\sigma=20$.

\item $\kappa_{\min}$  : the threshold value of the cornerness measure to define the set $Corners$. In all the experiments presented, we fix $\kappa_{\min}=0.5$.

\end{enumerate}
\end{document}